\documentstyle[aps,twocolumn]{revtex}
\renewcommand{\vec}[1]{\mbox{\boldmath$#1$}}
\begin{document}
\twocolumn[\hsize\textwidth\columnwidth\hsize\csname
@twocolumnfalse\endcsname

\title{
Vertically coupled double quantum dots in magnetic fields
}
\author{Hiroshi Imamura}
\address{CREST and Institute for Materials Research, Tohoku University, Sendai
980-8577, Japan}
\author{Peter A. Maksym}
\address{Department of Physics and Astronomy,
University of Leicester, Leicester LE1 7RH, UK.}
\author{Hideo Aoki}
\address{Department of Physics, University of Tokyo, 
Hongo, Tokyo 113-0033, Japan}
\maketitle

\begin{abstract}
Ground-state and excited-state properties of vertically coupled double quantum
dots are studied by exact diagonalization. 
Magic-number total angular momenta that minimize the total energy 
are found to reflect a crossover between 
electron configurations dominated by intra-layer 
correlation and ones dominated by inter-layer correlation. The position of 
the crossover is governed by the strength of the inter-layer 
electron tunneling and magnetic field.  
The magic numbers should have an observable effect on 
the far infra-red optical absorption spectrum, since 
Kohn's theorem does not hold when 
the confinement potential is different for two dots. This 
is indeed confirmed here from a numerical calculation 
that includes Landau level mixing. Our results take full account of 
the effect of
spin degrees of freedom. A key feature is that the total spin, $S$, 
of the system and the magic-number angular momentum are intimately 
linked because of strong electron correlation.
Thus $S$ jumps hand in hand with the 
total angular momentum as the magnetic field is varied.  
One important consequence of this is that the spin blockade 
(an inhibition of single-electron tunneling) should occur in some
magnetic field regions because of a spin selection rule.
Owing to the flexibility arising from the presence of 
both intra-layer and inter-layer correlations, the spin blockade 
is easier to realize in double dots than in single dots. 

\end{abstract}


\vskip2pc]

\section{Introduction}

Vertically coupled quantum dots have recently attracted much interest, 
since they open up the possibility of manipulating electron 
configurations in a three-dimensional space.  
Quantum dots are generally thought of as 
`artificial atoms', in which electrons are 
confined by an artificial electrostatic potential instead of
being attracted to a nucleus.
Numerical studies of few electron systems confined in single quantum dots
have shown that the angular momentum and spin of the ground state in 
strong magnetic fields belong to a special series values called magic 
numbers. For single quantum dots the interval of magic number angular
momenta has a one-to-one correspondence with the symmetry of charge
correlation. This has been explained by 
an `electron-molecule' picture by one of the 
present authors~\cite{map_eckart}.  Namely, the Coulomb 
repulsion forces the electrons confined in a dot 
to take a definite molecular configuration (triangle 
for three electrons, square for four, etc) 
when the electron correlation is strong~\cite{map_eckart,hima3}.  
If we use a similar picture for vertically coupled dots we can 
envisage the ground states in strong magnetic fields 
as three-dimensional electron molecules.  

On the experimental side, 
recent advances in nano-lithography and thin-film processing make
it possible to fabricate vertically coupled multiple quantum dots, 
where two-dimensional (2D) electrons are confined 
within an area smaller than 1 $\mu$m across~\cite{merkt1}.
Observable effects in the atom-like physics of dots 
have been detected from 
measurements of the tunneling current or capacitance.  
In the low magnetic field
regime ($B\leq$ 2T), Tarucha et al have indeed found from 
capacitance spectroscopy that Hund's rule and shell
structures appear in the spectrum~\cite{tarucha}.  
As the magnetic field becomes stronger, however, the single-electron
energy levels change to ones like those of a 2D harmonic oscillator to Landau
levels, where levels with different angular momenta are degenerate,
and electron correlation should play an essential role.  
Ashoori et al and Wagner et al have looked at 
the addition spectrum (energies required for adding one additional
electron) for single dots in magnetic fields, where 
a total-spin transition caused by the Coulomb
interaction is expected~\cite{ashoori1,ashoori2,wagner}.

This reminds us of the fractional quantum Hall
effect (FQHE) in the bulk 2D electron system, a manifestation of 
strong electron correlation in high magnetic
fields~\cite{prange,chakraborty,sarma}, where 
the total spin is a sensitive function of the 
density of electrons in the Landau levels. 
This is an effect of electron correlation, 
i.e., the way the electrons are correlated is strongly 
dominated by the total spin, while the Zeeman energy, 
which is few percent of the typical Coulomb energy, 
has only a minor effect.   

The fractional quantum Hall effect in double 
layers \cite{eisenstein,moon,nakajima,nakaCF,suen,yoshioka2} 
has recently been under intense study. In this case 
the additional degree of freedom arising from the double layer
(a pseudo-spin that labels the layers) enriches the physics. 
A central issue in these systems is the interplay of electron 
correlation and inter-layer electron tunneling.  
The competition between these effects makes the quantum Hall
state evolve continuously from a correlation-dominated (two-component) state
to a tunneling-dominated (single-component) state within the
quantum Hall regime. If we include spin degrees of freedom, 
FQH states specific to double-layer systems of 
electrons with certain specific spins indeed appear \cite{nakajima}.
Thus, vertically-coupled quantum dots are 
also intriguing from the viewpoint of what happens if 
we laterally confine a double-layer FQH system. 

In the present paper we investigate the physics of double dots in 
magnetic fields. 
While we have studied the magic numbers
and far infrared (FIR) absorption in our previous publications~\cite{hima1,hima2}, 
we have assumed there that the system is
fully spin polarized 
(except for our paper on the spin blockade, as 
recapitulated here in section \ref{sec:spin_blockade}).   
We have also adopted the lowest Landau level approximation there.  
These assumptions, however, are justified only for
$B\rightarrow\infty$ limit. 

One of our most important findings is that
the total spin of the ground state of the dots changes
wildly as the magnetic field varies, where 
the ground state is spin-{\it un}polarized even for a magnetic field 
as large as $B=4$ T in typical conditions. 
We show that this is a manifestation of 
electron correlation rather than the spin Zeeman effect.  
Namely, the electron correlation energy
is drastically affected by the total spin state 
even when we ignore the Zeeman term, so that it is the 
former that determines the spin.  
A similar phenomenon occurs in single dots and is explained by
the electron-molecule theory that takes care 
of spin quantum numbers~\cite{map_eckart,hima3}.
The wild change in the spin can cause  
single-electron tunneling to be blocked in some magnetic field regimes
due to spin selection rules as we have proposed recently~\cite{hima4}.  
Evidence that this effect is caused by electron correlation 
is the occurrence of a re-entrant non-blocked region  
which is hard to explain by Zeeman-energy considerations.

We have also investigated the effect of the
difference in the curvature of the 
confinement potential in the two layers. This can be
regarded as a pseudo-spin Zeeman energy and 
a novel charge correlation
caused by the pseudo-spin Zeeman energy is found to appear, which
causes dependence of the magic numbers on the strength of the inter-dot
electron tunneling and the layer separation. The upper branch in the 
absorption spectrum that approaches the cyclotron 
frequency for large $B$ has been obtained from a calculation that
includes higher Landau levels. We propose that the magic
numbers should have an observable effect on both the 
higher- and lower-branches of the FIR spectrum.  
Specifically, while the generalized Kohn theorem \cite{bery,map_prl} 
states that the electron correlation 
cannot affect the optical absorption
spectra in single dots with parabolic confinement, 
this is no longer the case with vertically coupled dots that 
have different confining potentials. 
We show that the absorption energy of the double dot should indeed exhibit 
discontinuities at the magnetic fields where the total angular
momentum or total spin change from one magic number to another.

Thus the purpose of 
this paper is two-fold: to include the spin degrees of freedom and 
to include the mixing of higher Landau levels, 
which enables us to calculate the real ground state.  
The magic numbers in double dots are found to depend on the strength
of the inter-dot tunneling and 
this determines how the total spin ($S$) 
changes hand in hand with the total angular momentum ($L$), which 
can be distinct from the link between $S$ and $L$ in single dots. 

The organization of this paper is as follows:
In Sec.~\ref{sec:model} the theoretical model of a vertically coupled
quantum dot is described. The physics of electron correlation in 
double dots and its relevance to the magic numbers is discussed in 
section Sec.~\ref{sec:ddot_magic}.  
The remaining sections deal with the experimental consequences of our 
results. All the results are summarized in Sec.~\ref{sec:conclusion}

%============================================================
%============================================================

\section{Vertically coupled quantum dots}
\label{sec:model}
Quantum dots are fabricated with a variety of
techniques.  The difference between these techniques lies in the way 
the lateral confinement potential of two-dimensional electron
system is created.
One method for obtaining lateral confinement is
to etch away a semiconductor sample 
to obtain mesas which contain a laterally confined two-dimensional
electron system.  The dot structure fabricated in this way is
called a ``deep-mesa-etched'' quantum dot (Fig.~\ref{fig:double_quantum_dot} (a)).
Another method is to deposit a metallic gate on top
of a heterostructure. When biased negative the gate will deplete 
electrons from the region underneath it thereby creating a quantum dot (Fig.~\ref{fig:double_quantum_dot} (b)).
Simple electrostatic considerations show that 
the potential felt by an electron in a mesa-etched sample 
is similar to that of a charged disk.  
The electrostatic potential of a charged disk can be 
found in terms of elliptic integrals and the bottom of the potential
is well approximated by a parabola~\cite{map_prl,johnson}.

Recent advances in semiconductor fabrication techniques have 
enabled fabrication of double dots in vertical, triple-barrier 
structures on submicron scales.  
Fig.~\ref{fig:double_quantum_dot}(a) actually depicts such a case: 
a gated field effect confined double dot and 
a deep-mesa-etched double dot with the source gate on top,
which has been studied by Tarucha et al~\cite{austing}.

As mentioned, we can assume that 
the confining potential of a quantum dot is 
parabolic.  Then the single electron states
are those of two dimensional harmonic oscillator.
When a perpendicular magnetic field 
$B$ is applied, the Hamiltonian for a single electron in a 
quantum dot is given by
\begin{equation}
  {\cal H} = \frac{1}{2m^{*}}\vec{\pi}^{2}
           + \frac{1}{2}m^{*}\omega_{0}^{2}r^{2}
           + \frac{1}{2}g^{*}\mu_{B} B \sigma_{z},
\end{equation}
where $m^{*}$ is the effective mass, $\omega_{0}$ the strength of
the parabolic confinement potential, $\mu_{_{B}}$ is the Bohr magneton, $g^{*}$ the effective
$g$-factor, $\frac{1}{2}\sigma_{z}$ the $z$-component of the spin of a single electron, and the canonical momentum
$\vec{\pi}$ is given by
\begin{equation}
  \vec{\pi} = - i \hbar \vec{\nabla} + e {\bf A},
\end{equation}
where the vector potential ${\bf A}$ satisfies 
$\vec{\nabla} \times {\bf A} = {\bf B}$.
For parabolic confinement, the Hamiltonian is similar to 
that for a free electron, so that 
the eigenstates $\phi_{n\ell}$, 
called Fock-Darwin states~\cite{fock,darwin},
are given by 
\begin{eqnarray}
  \phi_{n\ell}({\bf r})
  =&&
  \sqrt{\frac{n!}{2\pi \lambda^{2}2^{|\ell|}\lambda^{2|\ell|}(n +
      |\ell|)!}}
  \nonumber\\
  &&\times
  r^{|\ell|} {\rm e}^{- i \ell \theta} 
  L_{n}^{|\ell|}\left( \frac{r^{2}}{2 \lambda^{2}}\right)
  {\rm exp}\left( -\frac{r^{2}}{4\lambda^{2}}\right).
\label{eq:fock_wf}
\end{eqnarray}
Here $n$ is the radial quantum number, $\ell$ the angular momentum 
quantum number, $L_{n}^{|\ell|}$ a Laguerre polynomial,
$\lambda = \sqrt{\hbar/m^{*}\Omega}$ is the effective magnetic length, 
$\Omega\equiv\sqrt{\omega_{c}^{2} + 4\omega_{0}^{2}}$ 
and $\omega_{c} = e B / m^{*}$ is the cyclotron
frequency. The Fock-Darwin states are 
ring-shaped with a radius $R\sim \lambda\sqrt{2(2n+|\ell|+1)}$.
The eigenenergies are given by

\begin{equation}
  E_{n\ell\sigma}(\Omega )
  = \frac{1}{2}(2n + 1 + |\ell|) \hbar \Omega
  - \frac{1}{2} \ell \hbar \omega_{c} +\frac{1}{2}g^{*}\mu_{_{B}} B \sigma_{z}.
\label{eq:e_fock}
\end{equation}

For double quantum dot systems, the single-electron states are split into
symmetric and antisymmetric states due to inter-layer tunneling (Fig. 2).
The strength of the tunneling is characterized by the
energy gap, $\Delta_{_{\rm SAS}}$, between symmetric and anti-symmetric
states.  
Although $\Delta_{_{\rm SAS}}$ depends ($\sim$ exponentially) on the layer
separation $(d)$, it also depends on the height of the barrier 
that separates the two layers. Experimentally, $d$ and 
$\Delta_{_{\rm SAS}}$ are independently adjustable.  
We can see that $d$ controls the difference between 
the intra-layer interaction, $\propto 1/r$, and 
the inter-layer interaction, $\propto 1/(r^2+d^2)^{1/2}$.   

The Hamiltonian for interacting electrons in a vertically coupled quantum dot,
\begin{equation}
  {\cal H} = {\cal H}_{0} + {\cal H}_{\rm t} + {\cal H}_{\rm C},
\label{eq:hamiltonian}
\end{equation}
comprises the 
single-electron part, ${\cal H}_{0}$, 
and the Coulomb interaction, ${\cal H}_{\rm C}$.
In second-quantized form 
with a Fock-Darwin basis we have 

\begin{equation}
  {\cal H}_{0} = \sum_{n}\sum_{\ell}\sum_{\sigma}\sum_{\alpha}
  \varepsilon_{n\ell \sigma\alpha}
  c_{n \ell \sigma\alpha}^{\dag}c_{n \ell \sigma\alpha},
\end{equation}

\begin{equation}
  {\cal H}_{\rm t} = -\frac{\Delta_{\rm
      SAS}}{2}\sum_{n}\sum_{\ell}\sum_{\sigma}
  \left(
    c_{n \ell \sigma +}^{\dag}c_{n \ell \sigma -}
    + c_{n \ell \sigma -}^{\dag}c_{n \ell \sigma +} \right),
\end{equation}

\begin{eqnarray}
  {\cal H}_{\rm c} &=&
  \frac{1}{2}\sum_{n_{1}\sim n_{4}}
  \sum_{\ell_{1}\sim\ell_{4}}\sum_{\sigma_{1}\sim\sigma_{4}}
  \sum_{\alpha_{1}\sim\alpha_{4}}\nonumber\\
  &&\langle n_{1} \ell_{1}\sigma_{1}\alpha_{1},
  n_{2} \ell_{2}\sigma_{2}\alpha_{2} |
  V({\bf r}_{1}-{\bf r}_{2})
  |n_{3} \ell_{3}\sigma_{3}\alpha_{3},
  n_{4} \ell_{4}\sigma_{4}\alpha_{4}\rangle\nonumber\\
  &&\times  c_{n_{1} \ell_{1}\sigma_{1}\alpha_{1}}^{\dag}
  c_{n_{2} \ell_{2}\sigma_{2}\alpha_{2}}^{\dag}
  c_{n_{4} \ell_{4}\sigma_{4}\alpha_{4}}
  c_{n_{3} \ell_{3}\sigma_{3}\alpha_{3}}.
\end{eqnarray}
Here $\sigma$ is the real spin, $\alpha= \pm$ is a pseudo-spin index
specifying the two layers, $c_{n \ell \sigma \alpha}^{\dag}, (c_{n \ell
  \sigma \alpha})$ are creation 
(annihilation) operators and $\epsilon$ is the dielectric constant
of the host material.
When the spin and pseudo-spin are included the energy of the 
single-particle states becomes
\begin{equation}
  \varepsilon_{n \ell \sigma \alpha}
  = 
  E_{n\ell\sigma}(\Omega_{\alpha}),
\end{equation}
where $E_{n\ell\sigma}$ is given by eq.(\ref{eq:e_fock}) and 
$\Omega_{\alpha} = (\omega_c^{2}+4\omega_{0\alpha}^{2})^{1/2}$ 
with $\omega_{0\alpha}$ the confining potential for the 
$\alpha$\/th layer.

Because the Coulomb interaction conserves the total angular
momentum the many-body 
Hamiltonian is diagonalized
numerically in each sector of the total angular momentum
space~\cite{map_prl,laughlin2,girvin,galejs,stone}.

%=============================================================================
%=============================================================================

\section{Magic-number states}
\label{sec:ddot_magic}
One of the most dramatic features of interacting two dimensional
electron systems confined in quantum dots subjected to magnetic fields
appears in their energy spectrum.
Numerical studies of finite single-layer systems
have shown that ground states only occur at
certain total angular momenta and total spins called ``magic numbers''. 
The reason why magic numbers appear is that the Coulomb part of the
total energy is not a smooth function of the total angular momentum, $L$, 
but has a series of cusps caused by electron correlation~\cite{map_prl}.
A larger $L$ corresponds to a larger 
spatial extent ($\sim \sqrt{L}$) of the charge density, which 
costs a higher confinement-potential energy 
(Fock-Darwin energy) while the Coulomb repulsion tends to be reduced. 
What the magic numbers are all about is that 
the competition of these effects is not monotonic.  
Indeed, the  magic numbers have a one-to-one correspondence with the
symmetry of charge correlation, which has turned out to be 
the same as the symmetry of
the classical equilibrium configuration of point charges for single dots.
If we further take account of the electron spins, we have  
to consider, in addition to the Zeeman energy, 
the competition among the states having various total
spins that determines how the electrons should be correlated.  

Magic number states also occur in double dots. For double dots, however, 
we have pseudo-spin as an additional degree of freedom, which makes the
sequence of magic numbers for double dots different from that of single
dots. We first look at how the sequence of magic numbers is modified by
the strength of inter-layer tunneling.

\subsection{Three-electron double dots}

For simplicity we start with three fully-spin-polarized electrons in a
double dot with the same confining potential for both layers, while
the effect of spin will be examined later. We plot the ground-state energy
(Fig.~\ref{fig:ddot_new_magic})
calculated numerically as a function of the total angular momentum, $L$, 
at $B=15$T for $\Delta_{\rm SAS}=0.2 (0.6)$ meV in the upper (lower) panel. 
The confinement energy is $\hbar \omega_{0} = 3.0$meV for both
layers and the layer separation $d=20$ nm.

Magic numbers can be 
identified from the positions of downward cusps. 
For $\Delta_{\rm SAS}=0.2$ meV 
we have a period of two ($L=5, 7, 9$)
for smaller $L$ 
followed by a period of three ($L=9, 12, \cdots$).
For a larger $\Delta_{\rm SAS}=0.6$ meV, by contrast, 
the period is three for all $L$,
as in the case of a single-dot containing three electrons.

In the right panel of Fig.~\ref{fig:ddot_new_magic} 
we show the average charge density, $\rho$, 
which is axially symmetric.  
Since we assume the same confinement potential for the two
dots, $\rho$ is the same for the two dots.
For $L=5$ (solid line) 
the density against the lateral distance $r$ from the 
center has a peak 
at the center along with a shoulder around 
11 nm (=$1.7 \lambda, \lambda$: effective magnetic length is 6.54 nm).
For $L=12$ belonging to the period of three (long dashed line), on the
other hand, the density is peaked at a finite $r$.

In order to identify the mechanism for the change of period 
in the magic numbers, a better measure of the electron correlation is 
the pair correlation function $P({\bf r}, {\bf r}_{0})$ 
which is proportional to 
the conditional probability of finding an electron at
position ${\bf r}$ given that there is one at position ${\bf r}_0$. 
This is shown in Fig.~\ref{fig:ddot_new_magic_pc}. 
We fixed one electron at the place 
where the charge density has a 
shoulder (or a maximum when the density is peaked at a finite $r$). 
We can immediately see that the ground-state electron configuration 
changes from the one dominated by inter-layer correlation 
to one dominated by intra-layer correlation. In particular,
the correlation for $L=5$ corresponds to a triangular ``electron 
molecule'' developed {\it across} the two layers, with  
one electron at the center of the lower layer while 
the other two are in the upper layer. In contrast, the 
triangular form develops within each layer for $L=9,12$.  
The $L=7$ state is an intermediate one where the upper layer has a single 
peak but the lower layer has two peaks.

%Three-dimensional plots of the pair correlation function are shown in
%Fig.~\ref{fig:cf_n3_d_same_all}, where the sphere represents the fixed 
%electron.  One can easily see how the electron
%correlation changes form one with two-fold symmetry (L=5 and 7) to one
%with three-fold symmetry (L=9 and 12) as $L$ increases. 

An important difference between a single dot and a double
dot is the way a magnetic field affects the plot of ground-state 
energy $E$ against $L$.  For single dots and strong fields such that 
states with $n=0$ and $\ell \ge 0$ are dominant, the Fock-Darwin 
energy in Eq.~\ref{eq:e_fock} becomes $\hbar(\Omega - \omega_c)\ell/2 +
\hbar\Omega /2$. The main effect of changing the magnetic field 
is then to change the coefficient of the term linear in $\ell$
so that once we obtain an $E-L$ plot for a 
specific value of $B$ we can predict how 
the ground-state $L$ will evolve with $B$.  
In other words, a series of cusps (magic numbers) in the $E-L$ 
plot will sequentially become the true ground state as $B$ is varied. 
For double dots a change in $B$ affects not only the linear term but 
also the ratio of the effective magnetic length $\lambda$ and the layer
separation $d$, or equivalently, the ratio of the intra- and 
inter-layer Coulomb interactions.
Thus we have to check which of the magic number states 
can become true ground states.

We have done this in 
Fig.~\ref{fig:gs_gl_n3_d_same}.   
Here we have also included the effect of spin degrees of freedom.  
So what we have plotted in the figure is the total spin as well as 
the total angular momentum of
the ground state of the three-electron double dot
against magnetic field for $\Delta_{_{\rm SAS}}=0.2(0.6)$ meV in the
left(right) panel.
One can find that the magic-number states for $(L=5,S=3/2)$ or
$(L=7,S=3/2)$ become the absolute ground state for $B=7.2 \sim 7.5$T
or $B=10.2\sim 11.0$ T, respectively for $\Delta_{_{\rm SAS}}=0.2$ meV.
On the other hand, the magic numbers $L=5,7$ 
are washed out when  $\Delta_{_{\rm SAS}}$ is increased to 0.6 meV.

\subsection{Four-electron double dots}

The crossover between dominant intra-layer correlation and 
dominant inter-layer correlation also exists in four-electron systems.  
The total angular momentum and the total spin of the ground state of
four electrons are shown in Fig.~\ref{fig:gs_gl_rho_n4_d_same}.  
The system is fully spin-polarized for B $\geq3.2$T, where the magic-number 
orbital angular momentum evolves as $L=4,6,8,10,14$.  
In this regime we can discuss the magic numbers without 
complications coming from real spins. 

The charge density in Fig.~\ref{fig:gs_gl_rho_n4_d_same} 
is single-peaked for $L$ up to 14.  
However, unlike the three-electron case, 
this does not imply that the correlation is developed within each layer.  
If we look at the pair correlation function in Fig.~\ref{fig:pc_n4_d_same}, 
we can in fact see that the correlation is developed across the 
two layers.  Since we have four electrons we have now 
a tetrahedral configuration 
(with two electrons in the upper layer, and the other two in the lower 
layer in a staggered position) for all the magic-number states.  

As the magnetic field increases, the period of magic numbers changes
from two ($L=6+2\times{\rm integer}$) to four ($L=6+4\times{\rm
integer}$) as shown in the left panel of
Fig.~\ref{fig:gs_gl_rho_n4_d_same}.
The absence of the magic number $L=12$ is strong evidence for this
change, because $L=12$ belongs only to the sequence $L=6+2\times{\rm
integer}$.
The change of the period of the magic numbers corresponds to a
crossover between dominant intra-layer correlation and dominant
inter-layer correlation by analogy with three electron systems,
where the period of the magic numbers changes from two ($L=5,7$) to
three ($L=9,12$) and $L=11$ is the missing magic number. 
The results for the pair correlation function in
Fig.~\ref{fig:pc_n4_d_same},
also show that inter-layer correlation develops as $L$
increases.

%%%For larger angular momenta a crossover to dominant  
%%%intra-layer correlation occurs, where there are four peaks per layer in 
%%%$P({\bf r}, {\bf r}_{0})$.  
%%%This is indeed seen from the result that the period of the magic
%%%numbers changes from two for smaller $L$ to four 
%%%for larger magnetic fields with magic $L\geq 10$.  
%%%There the sequence $L=6+2\times {\rm integer}$ $(L=4,6,8)$ 
%%%corresponds to a two-by-two configuration 
%%%while $L=6+4\times {\rm  integer}$ $(L=6,10,14)$, 
%%%corresponds to a four-by-four configuration in each layer.  
%If we see the result for the pair correlation function 
%one can see that the shape of pair correlation indeed 
%changes from two-by-two to four-by-four.
%pam: this is not clear.
%%HA: This is NOT AT ALL clear ???
%pam: this is still not clear. Part of the problem is that the 
%competing 2x2 configuration is allowed to occur at the same L values 
%as the 4x4 configuration. Presumably the 4x4 configuration dominates 
%when L is very large but figure 7 does not show results for very 
%large L. The paper would be much stronger if a clear 4x4 configuration 
%could be shown. Do you have results for larger L that could 
%be shown in place of the L=14 result?

\subsection{Crossover between intra- and inter-layer correlations}
Intuitively the change in the correlation exemplified above for three 
and four electrons can be understood by considering the total energy.
If the total angular momentum $L$ is decreased at fixed magnetic field
the lateral spatial extent ($\propto \sqrt{L}$) of the wave function
is compressed, so that the mean electron separation, $a$, 
becomes comparable with the vertical separation, $d$, 
of the layers.  
When the two layers are close enough, $d \ll a$, the system 
is effectively one-layer, so that all the electrons are 
correlated (with a square configuration for the four-electron
case) within each layer.  
In the opposite case of separated layers, $d \gg a$, 
the electrons in a, say, four-electron system should 
tend to minimize the energy by separating 
into two in the upper layer and another 
two in the lower ending up with a tetrahedral configuration 
across the layers.  
This is because the typical intra-layer Coulomb repulsion 
($\propto 1/a$) 
becomes significantly greater than that of the inter-layer Coulomb interaction 
($\propto 1/\sqrt{a^2+d^2}$), so that 
the electrons will avoid each other 
as far as possible by `sidetracking', i.e., by developing 
inter-layer correlations.  
Such staggered configurations 
have to involve the mixing of symmetric and antisymmetric 
states across the two dots via the inter-layer tunneling, 
and costs an energy $\Delta_{\rm SAS}$.
This explains why the new magic numbers tend to emerge for smaller 
$\Delta_{\rm SAS}$.  
Thus we can call the change in the nature of the electron correlation 
as a crossover from a dominant intra-layer correlation to 
a dominant inter-layer correlation. Further, if $L$ is increased $a$ 
becomes large and the inter- and intra-layer interactions become 
similar. Then the system is again effectively one layer and this explains 
the tendency for the new magic numbers to occur for smaller $L$.
These qualitative arguments apply 
for any value of the magnetic field but the magnetic field determines 
the overall length scale and hence the $L$ value of the true ground state.

%In the above discussion we have fixed the magnitude 
%of the magnetic field $B$ for simplicity.  
%We have to vary $B$ 
%if we want to vary $L$ for the true ground state.  
%pam: I think that the discussion of jumps in R should either be 
%expanded or deleted. We have not shown any evidence that the jumps 
%occur in double dots so it is difficult for the reader to accept this 
%poin without further discussion. In addition, we do not know how the 
%average value of R behaves as B changes, ie are the jumps superimposed 
%on a constant value or a a curve?
%In this case the spatial extent $R$ of the wave function is 
%a function of $B$, but $R$ jumps as one magic $L$-value 
%goes over to another, so the argument still holds qualitatively.

% A second revision starts here -------------------------------------
%%%-------------------------
%%% pseudo spin Zeeman
%%%-------------------------
\subsection{Effect of the pseudo-spin Zeeman energy}
\label{sec:pseudo}
When the two layers have different confinement potentials, the
difference plays the role of a Zeeman energy for the
pseudo-spin~\cite{hima1,palacios},

\begin{equation}
  \varepsilon_{n \ell \sigma +}
  - \varepsilon_{n \ell \sigma -}
  = \frac{1}{2}(2n + 1 + |\ell|) \hbar (\Omega_+ - \Omega_-),
\end{equation}
where 
$\Omega_{\pm} \equiv (\omega_{c}^{2} + 4\omega_{0 \pm}^{2})^{1/2}$. 
The layer having the smaller confinement
potential tends to accumulate charge 
and this causes a charge density imbalance between the two layers as shown in
Fig.~\ref{fig:gs_gl_rho_n4_d_diff}
for a ten percent difference in 
$\hbar\omega_{0+}=3.0$ meV, $\hbar\omega_{0-}=3.3$ meV~\cite{hima1}.
For four-electron systems, the Zeeman
effect for pseudo-spin leads to a novel magic number which 
corresponds to 
tetrahedral-like charge correlations~\cite{hima2}. 

Magic angular momenta and the total spin in this case are displayed in 
Fig.~\ref{fig:gs_gl_rho_n4_d_diff}.  We can see that the real spins are 
fully polarized for 
$B\geq4$ T.  
Let us discuss this region in order to focus on the pseudo-spin.  
There the magic numbers are $L=6,9,10,12,14$, and 
we notice that the magic numbers $L=9,12$ 
do not appear in system without the pseudo-spin Zeeman energy. In particular,
$L=9$ belongs to neither the sequence 
$L=6+2\times {\rm integer}$ nor $L=6+4\times {\rm integer}$. Instead it 
belongs to $L=6+3\times {\rm integer}$, which 
implies a three-fold symmetry in the charge correlation.  

%pam: I don't understand what is meant by average charge density. The 
%figure shows upper and lower layer charge densities. What has been 
%averaged?
The average charge
densities in the ground states are shown in 
Fig.~\ref{fig:gs_gl_rho_n4_d_diff} for both upper and lower layers.
The densities for $L=6,9,12$, which belong to the sequence $L=6+3\times
{\rm integer}$, are peaked at the center.  
If we look at the pair correlation function in Fig.~\ref{fig:pc_n4_d_diff}, 
the $L=9$ state is indeed seen to have a trigonal cone 
(three-by-one) configuration
(while the $L=6$ state may be thought of as 
a mixture of two-fold and trigonal configurations).  
The $L=12$ state belongs to both 
$L=6+3\times{\rm integer}$ and $L=6+2\times{\rm integer}$ sequences, 
so that its correlation is a mixture of conical (three-by-one) 
and tetrahedral (two-by-two) symmetries.

It would be interesting to compare our results with the phase
diagram for the bulk bilayer 
FQH system obtained experimentally~\cite{boebinger}, or
theoretically~\cite{macdonald}.  
The picture in the bulk is that the quantum Hall state 
evolves, as $\Delta_{\rm SAS}$ is decreased, 
from a tunneling-dominated state to a correlation-dominated one.  
The former is a fully occupied symmetric state (one-component state), while 
the latter is an inter-layer correlated Laughlin's liquid, 
$\Psi_{331}$ (two-component).  

To compare these with the dot states, however, is
not straightforward.  The bulk phase diagram
is drawn against two dimensionless quantities, $d/\ell_B$ and 
$\Delta_{\rm SAS}/(e^2/\epsilon \ell_B)$, 
where $\ell_B\equiv (\hbar /m^*\omega_c)^{1/2}$ is the magnetic length.  
Because of the confining potential, the relevant 
length scale for dots becomes the effective magnetic length
$\lambda$ with $\lambda^2 = \hbar /m^* (\omega_{\rm c}^2 +
4\omega_0^2)^{1/2}$.
With the parameters we have used $\lambda = (0.91\sim 0.97)\ell_B$
for $B=(5\sim 10)$T.  
This yields $e^2/\epsilon \lambda = 14.7$ meV for $B=10$T, so that 
$\Delta_{\rm SAS}/(e^2/\epsilon \lambda) = 0.01 \sim 0.04$ 
for the double dots considered here. 
The Landau level filling, $\nu$, which is 
usually defined as
\[
\nu = N(N-1)/2L
\]
for a dot having $N$ electrons 
ranges from $\nu=3/5$ for $L=5$ to $\nu=1/4$ for $L=12$ for $N=3$. 
Although there is some attempt at extending the bulk phase 
diagram from the usually studied $\nu=1$ case to 
$\nu=1/$(odd integer) \cite{nakajima}, 
wide variation in $\nu$ and the fact that a change in $B$ also 
changes $\lambda$ complicate the comparison.  
%Comment by HA ---  No mention of maximum density droplet?

%==============================================================================
%==============================================================================

\section{FIR Absorption}
\label{sec:ddot_fir}
Now we move on to the FIR absorption spectrum.
In a single dot with a parabolic confinement potential, 
the electron-electron interaction cannot affect the 
FIR absorption.  This is dictated by 
Kohn's theorem~\cite{kohn}, originally conceived for translationally 
symmetric systems and later generalized to 2D dots with parabolic 
confinement in magnetic fields \cite{bery,map_prl}.  
Namely, long-wavelength electromagnetic radiations with electric 
field ${\bf E}$ couples to the dot via the dipole interaction, 
\begin{equation}
{\cal H}^{\prime}
  = \sum_{i = 1}^{N} e {\bf E} \cdot {\bf r}_i,
\end{equation}
which depends only on the center-of-mass coordinate. In a single dot 
with parabolic confinement the Hamiltonian separates into 
the center-of-mass and relative (interaction) parts even in the
presence of magnetic fields, and the latter
is irrelevant to optical transitions. 
Recent optical measurements of
quantum dots indeed exhibit absorption frequencies that are
independent of the number of electrons and well fitted 
to the single-electron absorption spectrum~\cite{meurer}.

In contrast, the separation of 
the center-of-mass and relative parts 
does not occur in vertically coupled dots having different
confinement potentials even when they are both parabolic.  This means 
that the Coulomb interaction can affect FIR absorption spectra.  

To quantify the effect we have calculated the 
FIR absorption spectrum of vertically coupled dots 
from the matrix element of the perturbation Hamiltonian,
$ {\cal H}^{\prime} $, between the ground state and all 
the excited states. 
Before discussing the results, we comment on the applicability of
this approach to 
real systems. One important question is the nature of the electric
field ${\bf E}$ in small samples.  
Several authors have questioned the relation between the 
applied electric field and the internal electric field in mesoscopic 
systems \cite{Perenboom,Cho,Keller} with the general conclusion that 
depolarization effects are important. Therefore we would have to 
calculate the internal electric field to obtain the absolute value
of the absorption coefficient. In addition, precise calculation of
the absorption spectrum would require us to take account of other
device properties that affect absorption, such as finite thickness of
the individual dots and deviations from a parabolic potential, about
which 
scant information is available. We therefore make the reasonable 
assumption that the internal electric field is uniform and discuss
only 
the absorption energy and the relative intensities of various
transitions. 
This should be sufficient for our purpose of demonstrating 
that the FIR absorption of vertically 
coupled dots is affected by the electron correlation.

The FIR absorption spectrum consists of two branches: the upper branch for
inter-Landau-level transitions, which approaches to the
cyclotron frequency in the $B\rightarrow \infty$ limit, and the lower
one for intra-Landau level transitions.
If we plot the position and intensity of the peaks in the FIR spectra
along with the ground state total angular
momentum $L$ and spin $S$ (Figs.~\ref{fig:fir_n3} and \ref{fig:fir_n4}),
we find a one-to-one correspondence between the magnetic fields at which 
the absorption line jumps and the magnetic fields at which the total
angular momentum and/or the total spin of the 
ground state changes from one magic-number state to another.
Thus the transitions should be directly observable in
the FIR absorption spectrum. 
The figure also shows that the absorption intensity 
($\propto$ square of the matrix element) is not monotonic. 
We have also displayed the FIR spectra for a four-electron system, 
which again exhibit similar jumps. The magnitudes of the jumps, 
obtained assuming a confining potential asymmetry of 10\%, is a few 
tenths of a milli electron-volt. We believe they should be observable 
and the effect could probably be enhanced by making a double dot 
with greater potential asymmetry.

%==============================================================================
%==============================================================================

\section{spin blockade}
\label{sec:spin_blockade}
We now turn to the possibility of a 
spin-blockade.  
It has been known for some time that the Coulomb blockade occurs in 
mesoscopic systems such as quantum dots. 
This is a combined effect of the discreteness of energy levels 
and the electron-electron interaction (charging energy).  
Weimann et al then suggested that, 
if the total spin differs by more than 1/2 in 
the ground states for $N$ and $(N-1)$ electrons, i.e., 
\begin{equation}
|S(N)-S(N-1)| > \frac{1}{2}
\label{eq:spin_blockade}
\end{equation}
the transport of an electron through the dot 
is blocked, and this should cause missing peaks 
in the conductance due to single electron tunneling at zero temperature. This is 
called the spin-blockade~\cite{weinmann_prl,tanaka}
%and has been studied 
%theoretically for weak interaction regimes.  
%When the one-electrons levels are well-defined for 
%weak interactions, we can apply 
%the Hund's coupling picture, in which electrons are 
%accommodated in one-electron states and the spins are aligned 
%within degenerate states. 
and has been studied theoretically in the case of no magnetic field, where we can apply the Hund's coupling picture~\cite{tarucha}.
However, we cannot satisfy the spin-blockade condition for circular quantum dots with parabolic confinement potential in 
this case, so that some modifications such as 
an anharmonicity in the confinement
potential~\cite{eto} or rectangular hard wall shape confinement potential~\cite{weinmann_prl} have to be introduced.  

Now our idea here is to exploit the manifestation of 
the electron correlation in the total spin in the present system. 
We have in fact seen that the 
electron correlation dominates the total spin ($S$)
%, where $\vec{S}^{2}=S(S+1)$, --- this should appear previously?
of the ground state in a peculiar manner, 
as shown in figures~\ref{fig:gs_gl_n3_d_same},
\ref{fig:gs_gl_rho_n4_d_same}, and
\ref{fig:gs_gl_rho_n4_d_diff}, 
which comes from the fact that the magic values of $L$ and $S$ are 
linked. This is in a sense no surprise, since the total spin dictates, 
through Pauli's exclusion principle, 
the way in which the electrons can correlate, 
while the magic numbers in $L$ come from the electron correlation 
as one of the authors has shown from the electron molecule picture for 
single layer quantum dots.  

The correlation effect in fact 
happens when the typical Coulomb energy is much greater than 
the single-electron level spacing.  
Effects of electron correlation on the spin states 
via Pauli's principle are known to occur in
ordinary correlated electron systems such as the Hubbard model, but
the present case is a peculiar manifestation in strong magnetic fields.

Here we propose to utilize this electron correlation
effect to realize a spin blockade.  
For that we can look at our results for $S$ 
to see if the spin-blockade condition, Eq.~\ref{eq:spin_blockade},
is satisfied in some regions of $B$.  
In addition, if there are low-lying excited states through 
which an electron can be transferred, the blockade will not 
occur except at the temperatures lower than the excitation energy.  
Thus we have to study the energies of excited states as well.  
We have plotted these along with $|S(4)-S(3)|$ for double
dots that contain three or four electrons with the
same parabolic confinement potential for both layer.  
The total angular momentum and the total spin of the ground state are
shown in the right panel of Fig.~\ref{fig:gs_gl_n3_d_same} for $N=3$, 
and in the left panel of Fig.~\ref{fig:gs_gl_rho_n4_d_same} for $N=4$.

The difference between the total spin of three and four electron
systems, shown in the bottom panel of Fig.~\ref{fig:exc_gs_d_n3_n4}, 
indicates that 
the spin-blockade condition, $|S(4)-S(3)| > \frac{1}{2}$, 
is satisfied for 2.0 $\leq B \leq 3.1$ T and $B=3.7$ T.  
In that region the excitation energies 
(top and middle panels of Fig.~\ref{fig:exc_gs_d_n3_n4}) 
for $N=3$ and $N=4$ are of the order of 0.05meV, which is 
unfortunately only 
half the typical experimental resolution $\simeq 0.1$meV.  
However, it is possible to enhance the excitation energy by tuning
$\hbar\omega_{0}$, $d$,  or $\Delta_{_{\rm SAS}}$. 
For instance, a double dot with $\hbar\omega_{0}=6.0$ meV, $d=16$ nm, and
$\Delta_{_{\rm SAS}}=1.2$ meV has the excitation energy of about 0.12 meV,
%pam: I have changed this. I think you meant an observable spin blockade not 
%an observable energy. Please check.
which should be large enough to cause an experimentally observable 
spin-blockade between $N=2$ and $N=3$~\cite{hima_spin_blockade}.
We also notice a level crossing between the second and the third 
excited states around $B=2.4$T for $N=3$, which should appear 
in the addition energy spectra.
%==============================================================================
%==============================================================================

\section{Conclusion}
\label{sec:conclusion}
We have studied double quantum dot systems by using
exact diagonalization including spin degrees of freedom and higher
landau levels.  
We have found novel magic numbers specific to double dots and shown that
they are related to the symmetry of charge correlations.  In addition, 
we have shown that we can change the charge correlation, total angular 
momentum, and total spin by varying the strength of the magnetic field $B$.
These changes have been shown to affect the optical absorption spectrum and
single electron tunneling.
For the optical absorption spectrum, we found that because of the
breakdown of Kohn's theorem, the spectra differ from that of
non-interacting system with splitting and jumps in the optical
absorption spectrum. The magnetic field where the spectrum has jumps 
is the field where the total angular momentum and/or the total spin
of the ground state shift from magic value to another. In the case of
single electron tunneling, we have shown that the 
tunneling is blocked because of the spin selection rule ---
the spin blockade, in some magnetic field regions.

\acknowledgements

We wish to thank Prof. Seigo Tarucha and Dr. David Guy Austing for a
number of illuminating discussions. 

%%\bibliographystyle{prsty}
%%\bibliography{ddot}

%\clearpage
\begin{figure}
  \caption{A deep-mesa etched (gated) vertically coupled quantum dot is
    shown schematically in a(b).
    }
  \label{fig:double_quantum_dot}
\end{figure}

%\clearpage
\begin{figure}
  \caption{
    Schematic single-electron
    wavefunctions in the direction normal to the layers in 
    a double quantum dot. 
    }
  \label{fig:double_well_wfunc_dots}
\end{figure}

%\clearpage
\begin{figure}
  \caption{
    Left: Ground-state energy against the total angular momentum,
    $L$, in vertically-coupled dots
    with three spin-polarized electrons for 
    $\Delta_{\rm SAS}=0.2(0.6)$ meV at $B=15$ T in the
    upper (lower)
    panel. The confinement energy 
    is $\hbar \omega_{0} = 3.0$ meV for both 
    layers, and the layer separation is $d = 20$ nm.
   Arrows indicate the positions of the cusps.  
   \newline
   Right:
   Charge density $(\rho)$ per electron for each layer 
   for $L= 5,7$ (belonging to period two) and $L=9,12$ (three) with
   $\Delta_{\rm SAS}=0.2$ meV.
    }
  \label{fig:ddot_new_magic}
\end{figure}

%\clearpage
\begin{figure}
  \caption{ a) : Contour plot of intra-layer (upper panels) and
    inter-layer (lower panels)
    pair correlation functions, $P({\bf r}, {\bf r}_{0})$, for $L=5,
    7, 9,$ and $12$.
    One electron (filled circle) is fixed at the place where the charge
    density has a maximum in the upper layer. 
    Thus the total charge is 1/2 (3/2) in the upper (lower), since we have 
    3/2 + 3/2 electrons while we fix one electron in the upper layer.  
    An area with a linear dimension of $12 \lambda = 78.5 {\rm nm} 
    ~(\lambda$: effective magnetic length) is displayed.
    The confinement energies, 
    the layer separation, $\Delta_{\rm SAS}$, and 
    $B=15$ T are the same as in the previous figure.
    \newline
    b): Bird's eye view plot of (a).  A white sphere indicates the
    position of the fixed electron.
    }
  \label{fig:ddot_new_magic_pc}
\end{figure}

%\clearpage
\begin{figure}
  \caption{The total angular momentum $L$ and the total spin $S$ of
    a double dot containing three electrons with confinement
    energy $\hbar\omega_{0+}=\hbar\omega_{0+}=3.0$ meV and layer
    separation $d=20$ nm. $\Delta_{_{\rm SAS}}=0.2(0.6)$ meV for
    the left(right) panel.
    }
  \label{fig:gs_gl_n3_d_same}
\end{figure}

%\clearpage
\begin{figure}
  \caption{
    Left: The total angular momentum $L$ and the total spin $S$ of
    a double dot containing four electrons with confinement
    energy $\hbar\omega_{0+}=\hbar\omega_{0-}=3.0$ meV.  The layer
    separation is $d=20$ nm and $\Delta_{_{\rm SAS}}=0.6$ meV.
    \newline
    Right: Charge density $(\rho)$ per electron for each layer for
    typical ground states in the fully spin-polarized
    region, $B\geq3.2$T.
    }
  \label{fig:gs_gl_rho_n4_d_same}
\end{figure}

%\clearpage
\begin{figure}
  \caption{a) : Contour plot of intra-layer (upper panels) and
    inter-layer (lower panels) pair correlation functions, $P({\bf r},
    {\bf r}_{0})$, 
    are displayed for each of the magic-$L$ states 
    in the true ground state for the four-electron double dot 
    shown in Fig.~\protect\ref{fig:gs_gl_rho_n4_d_same} 
    for appropriate values of $B$ 
    (in contrast to Fig.4 where $B$ is fixed).  
    One electron (white sphere) is fixed in the upper layer at
    the place where the charge density has a maximum. 
    Thus the total charge is 1 (2) in the upper (lower) layer, since we have 
    2 + 2 electrons while we fix one electron in the upper layer.  
    An area with the linear dimension of $12 \lambda ~(\lambda$: 
    effective magnetic length) is displayed.
    The confinement energies, 
    the layer separation $d$ and  $\Delta_{\rm SAS}$, are the same as in
    Fig.~\protect\ref{fig:gs_gl_rho_n4_d_same}.
    \newline
    b): Bird's eye view plot of (a).  A white sphere indicates the
    position of the fixed electron.
    }
  \label{fig:pc_n4_d_same}
\end{figure}

%\clearpage
\begin{figure}
  \caption{
    Left: The total angular momentum $L$ and the total spin $S$ of
    a double dot containing four electrons with confinement
    energy $\hbar\omega_{0+}=3.0$ meV and $\hbar\omega_{0-}=3.3$ meV.
    The layer separation is $d=20$ nm and $\Delta_{_{\rm SAS}}=0.6$ meV.
    \newline
    Right: Average charge densities $(\rho)$ per electron for 
    upper (lower) layer in
    the ground states are displayed in the upper (lower) panel 
    for $L \geq 6$, for which the spins are fully polarized. 
    }
  \label{fig:gs_gl_rho_n4_d_diff}
\end{figure}

%\clearpage
\begin{figure}
  \caption{
    Contour plot of intra-layer (upper panels) and inter-layer (lower)
    pair correlation functions, $P({\bf r}, {\bf r}_{0})$, for the
    same ground states as shown in Fig.~\protect\ref{fig:gs_gl_rho_n4_d_diff}.
    One electron (white sphere) is fixed in the upper layer at the
    point where the charge density has a maximum. 
    An area with the linear dimension of $12 \lambda$ is displayed.
    The confinement energies, 
    the layer separation $d$, $\Delta_{\rm SAS}$, and the magnetic
    field $B$ are the same as in
    Fig.~\protect\ref{fig:gs_gl_rho_n4_d_diff}.
    \newline
    b): Bird's eye view plot of (a).  A white sphere indicates the
    position of the fixed electron.
    }
  \label{fig:pc_n4_d_diff}
\end{figure}

%\clearpage
\begin{figure}
  \caption{ FIR absorption spectrum (top panel), the total angular
    momentum (middle), and the total spin (bottom)    of vertically
    coupled dots for $N=3$ electrons. $\hbar \omega_{0+} = 3.0$ meV, $\hbar
    \omega_{0-} = 3.3$ meV,     the layer separation $d = 20$ nm and
    $\Delta_{\rm SAS} = 0.6$ meV.    The position of each filled
    circle gives the energy of the    transition while the size of the
    circle represents the relative     intensity of absorption.
    Vertical dashed lines are guides to the eye. 
    }
  \label{fig:fir_n3}
\end{figure}

%\clearpage
\begin{figure}
  \caption{ FIR absorption spectrum (top panel), the total angular
    momentum (middle), and the total spin (bottom)    of vertically
    coupled dots for $N=4$ electrons is shown in the
    left(right) panel.     $\hbar \omega_{0+} = 3.0$ meV, $\hbar
    \omega_{0-} = 3.3$ meV,     the layer separation $d = 20$ nm and
    $\Delta_{\rm SAS} = 0.6$ meV.    The position of each filled
    circle gives the energy of the    transition while the size of the
    circle represents the relative     intensity of absorption.
    Vertical dashed lines are guides to the eye. 
    }
  \label{fig:fir_n4}
\end{figure}

%\clearpage
\begin{figure}
  \caption{
    Top(middle): Low-lying excitation energies for $N=4(N=3)$ double
    dots.\newline
    Bottom:  The absolute value of the difference, $|S(4)-S(3)|$,
    in the total spin for $N=4$ and $N=3$ double dots.\newline
    $\hbar \omega_{0+} = \hbar\omega_{0-} = 3.0$ meV, 
    the layer separation $d = 20$ nm and $\Delta_{\rm SAS} = 0.6$ meV.
    }
  \label{fig:exc_gs_d_n3_n4}
\end{figure}
\end{document}